\documentclass[letterpaper,oneside]{SprMixLaTeX}

\usepackage[sectionbib,longnamesfirst]{natbib}
\bibpunct{(}{)}{;}{a}{}{,}
\renewcommand{\cite}{\citep}

\usepackage{makeidx}         
\usepackage{graphicx,marvosym}        
\usepackage{multicol}        
\usepackage[bottom]{footmisc}

\usepackage{eucal,url}
\usepackage{graphics}

\usepackage{array}
\usepackage{latexsym}
\usepackage{amsmath}
\usepackage{amssymb}
\usepackage{color}
\usepackage{comment}
\usepackage[latin1]{inputenc}

\usepackage{algorithmic}
\usepackage{tabularx}
\usepackage{pbox}

\usepackage{psfrag}

\usepackage{verbatim}
\usepackage{import}

\usepackage{lscape}
\usepackage{subfigure}
\usepackage{rotating}

\usepackage{epstopdf}

\usepackage[ruled,algochapter]{algorithm2e}

\ifx\newtheorem\undefined
  \usepackage{amsthm}
\fi

\newlength{\defsep}
\makeindex             

\begin{document}

\frontmatter

\mainmatter

\definecolor{myColor}{rgb}{0.4,0.3,0.0}

\newcommand{\colorBox}[1]{
\setlength{\fboxrule}{1pt}
\fcolorbox{myColor}{white}{#1}
\setlength{\fboxrule}{0.5pt}
                       }


\title{Generating functionals for guided self-organization}

\author{Claudius Gros}

\institute{Institute for Theoretical Physics, Goethe University Frankfurt,
           Germany}

\maketitle

\vspace{0.7 cm}

\section{Controlling Complex Systems}
\label{section_Controlling_Complex_Systems}

One may take it as a running joke, that complex 
systems are complex since they are complex. It is 
however important to realize, this being said, that complex
systems come in a large varieties, and in many complexity
classes, ranging from relatively simple to extraordinary complex.
One may distinguish in this context between {\sl classical}
and {\sl modern} complex system theory. In the classical approach
one would typically study a standardized model, like the
Lorentz model or the logistic map, being described usually by 
maximally a handful of variables and parameters \cite{gros2008complex}.
Many real-world systems are however characterized by a very large number
of variables and control parameters, especially when it comes to
biological and cognitive systems. It has been noted, in this context,
that scientific progress may generically be dealing with complexity
barriers of various severities, in far reaching areas like
medicine and meteorology \cite{gros2012pushing}, when 
researching real-world natural or biological complex systems.

Generically, a complex system may be described by a set of
first-order differential equations (or maps), like
\begin{equation}
\dot x_i \ =\ f_i(x_1,x_2,\dots|\gamma_1,\gamma_2,\dots)~,
\label{dot_x_i}
\end{equation}
where the $\{x_i\}$ are the primary dynamical variables and
the $\{\gamma_j\}$ the set of control parameters. Modern 
complex system theory has often to deal with the situation where
the phase space of dynamical variables and parameters are both
high dimensional. Everything in the macroscopic world, f.i.\ the
brain, can be described by an appropriate set of equations of 
motion, like (\ref{dot_x_i}), and we are hence confronted with 
two types of control problems:
\begin{description}
\item{--} How do we derive governing equations of type (\ref{dot_x_i})?
\item{--} Given a set of equations of motion, like (\ref{dot_x_i}),
          how do we investigate its properties and understand the 
          resulting behavior as a function of the control parameters?
\end{description}
At its core, we are interested here in how to generically control,
in general terms, a complex and self-organizing system. A range of 
complementary approaches are commonly used in order to alleviate 
the control problem, we discuss here some of the most
prominent (non mutually exclusive) approaches.
\begin{itemize}
\item {\sl Delegation to Evolution}\newline
   One is often interested, especially in biology and in
   the neurosciences, in biologically realistic models and
   simulations \cite{markram2006blue}. In this case both the 
   functional form and the
   parametric dependences are taken from experiment. One may then
   expect, thanks to Darwinian selection, that the such constructed 
   dynamical equation should exhibit meaningful behavior, replicating 
   observations.
\item {\sl Exploring Phase Space}\newline
   A complete understanding would correspond, within dynamical system
   theory, to a full control of both the qualitative behavior of the
   flow in phase space and of its dependency on the control parameters.
   Achieving this kind of complete control is clearly very desirable, 
   but often extremely hard to achieve when dealing with large numbers
   of dynamical variables and control parameters, the typical situation
   in modern complex system theory. The exploration of phase space, 
   typically through a combination of analytical and numerical 
   investigations, is in any case an indispensable tool, even when 
   only a small fraction of the overall phase space volume can be probed.
\item {\sl Classical Control Theory}\newline
   Classical control theory deals with the objective to control a 
   real-world system, like a rocket, such that a desired behavior 
   is optimally achieved, in the wake of noise both in the sensor 
   readings and in the action effectiveness \cite{leigh2004control}. 
   Classical control theory is of widespread use in engineering and
   for robot control \cite{de1996theory}. Our present discussion 
   deals however with the general control of working regimens of a 
   self-organizing complex system; if we knew what the system 
   is supposed to do, we would be done.
\item {\sl Diffusive Control}\newline
   Neuromodulators \cite{marder2012neuromodulation}, like dopamine, 
   serotonin, choline, norepinephrine, neuropeptides and neurohormons,
   act in the brain as messengers of a diffusive control system 
   \cite{gros2010cognition,gros2012emotional}, controlling intrinsic and synaptic 
   properties like neural gain and threshold, or synaptic plasticity.
   Diffusive control is needed to stabilize a desired working
   regime, a process also denoted as metalearning \cite{doya2002metalearning},
   and to switch between different working regimes in order to
   achieve behavioral flexibility \cite{arnsten2012neuromodulation}.
   Diffusive control is a very general strategy for controlling
   a complex system.
\item {\sl Generating Functionals}\newline
   This is the subject we will develop here. One can achieve an
   improved understanding when considering classes of dynamical
   systems derived from superordinated functionals. In this case
   the equations of motion are not given a priori, but derived
   from a generating principle. Here we will detail out how this
   approach leads to an alleviation of the control problem.
\end{itemize}
One needs to recall, coming back to the introductory statement,
that there is no one-size-fits-all method for controlling
complex systems \cite{frei2011advances}, as there are many 
kinds and varieties of complex
systems. Here we will consider primarily systems made up of
a potentially large number of similar functional units, as
typical for neural networks. A related aspect of the generic 
control problem discussed above regards, in this
context, the stability of a default working regime with
regard to external influences and statistical fluctuations
\cite{clarke2007stability}. This is particularly important in
functional complex systems, such as an ecosystem
\cite{holling1973resilience,may2001stability}, or
cognitive architectures, the subject of our interest here.

\section{Guiding Self-Organization}
\label{section_Guiding_Self_Organization}

There is no strict scientific definition of what self-organization 
means or implies. It is however generally accepted to consider 
processes as self-organizing when a rich and structured dynamics 
results from a set of relatively simple evolution rules. The term
self-organization is of widespread use \cite{haken2006information}, 
ranging from classical non-equilibrium physical \cite{nicolis1989physics}
and biological \cite{camazine2003self} systems
to the assembly of complex macromolecules \cite{lehn2002toward};
it is quite generally accepted that the foundations of
life are based on self-organizing principles 
\cite{kauffman1993origins}. The brain in particular, 
possibly the most complex object presently known to humanity, 
is expected to result from a plethora of intertwined
self-organizing processes \cite{kelso1995dynamic},
ranging from self-organized cognitive functions
\cite{kohonen1988self} to self-organized critical states
\cite{bak1999nature,chialvo2010emergent}.

Self-organization is, per se, content free, having no
semantic relevancy. The stars in a rotating galaxy, to 
give an example, may spontaneously organize into a
set of distinct density waves, known as the arms of a
spiral galaxy. Even though pretty to the eye, the
spiral arms of the Milky Way do not serve any purpose;
self-organization is in this case just a byproduct of 
Newton's law. The situation is however generically
distinct for biological settings, or for man-made
systems, where functionality is the key
objective.

The design of functionality is of course a standard
objective for the vast majority of man-made systems, and
contrasts with the absence of functionality of natural
phenomena. Here we are interested in self-organizing
processes which are neither fully designed nor without any
objective. There is a middleway, which has been denoted 
{\sl ``targeted self-organization''} \cite{gros2008complex} or,
alternatively, {\sl ``guided self-organization''}
\cite{prokopenko2009guided,martius2010taming}.
\begin{center}
\colorBox{
  \parbox{0.20\textwidth}{
     \centerline{\phantom{\Large$|$}designed\phantom{\Large$|$}}
        }} $\longrightarrow$
\colorBox{
  \parbox{0.20\textwidth}{
     \centerline{\phantom{\Large$|$}guided\phantom{\Large$|$}}
        }} $\longrightarrow$
\colorBox{
  \parbox{0.20\textwidth}{
     \centerline{\phantom{\Large$|$}natural\phantom{\Large$|$}}
        }}
\end{center}
For a designed system the functionality is specified in detail
in order to achieve optimal performance for a given task.
The target for a self-organizing process is however presumed
to be a generic principle, often based on information theoretical
considerations, with the actual functionality arising indirectly 
through self-organizing processes. Targeted and guided
self-organization are essentially identical terms, with
guided self-organization having a somewhat broader breath.
One could guide, for example, a dynamical system by restricting
its flow to a certain region in phase space, allowing for
an otherwise unrestricted development within this bounded
area of phase space. Here we will neglect the differences
in connotation between targeted and guided self-organization 
and use both terms interchangeably.

Let us come back at this point to the general formulation 
of a complex dynamical system through a set of 
parameterized first-order differential equations,
as given by (\ref{dot_x_i}). The distinction between
a parameter $\gamma_j$ and a primary dynamical variable $x_k(t)$
is a question of time scales.
\begin{center}
\fbox{\parbox{6cm}{\centerline{
$\displaystyle
\left.
\begin{array}{rcl}
\dot x_k &:& \mathrm{fast} \\ 
\dot \gamma_k &:& \mathrm{slow} 
\end{array}
\right\} \quad \mathrm{time\ evolution}
$}
}}
\end{center}
The flow $(x_1(t),x_2(t),\dots)$
of the primary dynamics is taking place in the 
slowly changing environment of parameter space, defining 
the adiabatic background. The slow adaption of parameters 
is what controls in the end the working regime of a 
dynamical system, and is also denoted sometimes as 
metalearning \cite{vilalta2002perspective}. Not all 
parameters can be involved in metalearning,
a small but finite set of core parameters
$\{\gamma'_j\}\in\{\gamma_k\}$ must be
constant and immutable,
$$
\dot\gamma'_j \ =\ 0~.
$$
This set of core parameters is what defines in the end the
system. One has achieved a dimensional reduction of the 
control problem if the number $|\{\gamma'_j\}|$ of core 
parameters is small. This is the aim of guided
self-organization, that a concise set of core parameters
controls the development and the dynamical properties
of a system, with quantitative tuning of the values of
the control parameters inducing  modifications of
the system's characteristics, both on a quantitative and 
a qualitative level.

\section{Generating Functionals}
\label{section_Generating_Functionals}

There are two principle venues on how to
express guiding principles for dynamical
systems, implicitly or explicitly. In analogy,
one can implement conservation laws in physics
by writing down directly appropriate equations of
motion, demonstrating that, e.g., energy is conserved.
In this case energy conservation is implicitly present
in the formulation of the dynamical system.  Alternatively 
one may consider directly a time independent
Lagrange function, a condition which explicitly
guarantees energy conservation for the respective
Lagrange equations of motion. Here we will concentrate
on the second approach, the explicit derivation of
equations of motion for targeted self-organization through
appropriate generating functionals.

The term generating functional has a wide range of
connotations in the sciences. The action functional 
in classical mechanics and quantum field theory is a 
prominent example from physics, the generating functional
$\sum_k p_kx^k$ for a distribution function 
$p_k$ (with $p_k\ge0$ and $\sum_kp_k=1$) 
another from information theory. In the neurosciences
it is custom to speak of objective functions
\cite{intrator1992objective,goodhill1997unifying}
instead of generating functionals.

As a first example we consider a simple energy functional
\begin{equation}
E(\{x_k\}) \ =\ 
\frac{\Gamma}{2}\sum_k\, x_k^2\, -\,
\frac{1}{2}\sum_{kl}\,y(x_k) w_{kl} y(x_l),
\label{eq_functional_E}
\end{equation}
which is suitable for a network of neurons with
membrane potential $x_k$ and firing rate $y(x_k)$.
Here $y(x)$ is the sigmoidal transfer function 
\begin{equation}
y(x) = \frac{1}{1+\mathrm{e}^{a(b-x)}}~,
\label{eq_sigmoidal}
\end{equation}
parameterized
by the gain $a$ and the threshold $b$. The $w_{ij}$
in (\ref{eq_functional_E}) will turn into the 
synaptic weights, as we will show lateron, and
$\Gamma$ into a relaxation rate. Concerning the
terminology, one could consider $E(\{x_k\})$ also
to be an energy function (instead of a functional), 
being a function of the individual $x_k$. 
Here we use the term energy functional,
for the functional dependence on the vector 
${\bf x}=(x_1,x_2,\dots)$ of membrane potentials.

For our second example we consider a general functional
based on the principle of polyhomeostasis \cite{markovic2010self}.
One speaks of a homeostatic feedback loop when a
target value for a single scalar quantity is to be
achieved. Life per se is based on homeostasis,
the concentrations of a plethora of biological relevant
substances, minerals and hormones need to be regulated,
together with a vast number of physical properties, like
the body temperature or the heart beating frequency.
Polyhomeostasis is, in contrast, typically necessary for
time allocation problems.

The problem of allocating time for various tasks constitutes 
the foundation of behavior. Every living being needs to 
decide how much time to spend, relatively, on vitally 
important behaviors, like foraging, resting, exploring 
or socializing. Maximizing only a single of the possible 
behavioral patterns would be counterproductive, only a
suitable mix of behaviors, as an average over time, is 
optimal. Mathematically this goal is equivalent to 
optimizing a distribution function, hence the term 
polyhomeostasis, in contrast to the case of homeostasis, 
corresponding to the optimization of a single
scalar quantity.

All a neuron can do, at any given moment, is to fire 
or not to fire, a typical time allocation problem. 
The generic functional
\begin{equation}
F[p] \ =\ \int p(y) f(p(y))) dy
\label{eq_functional_poly}
\end{equation}
of the firing rate distribution
\begin{equation}
p(y) \ =\  \frac{1}{T}\int_0^T \delta(y-y(t-\tau))\,d\tau
\label{eq_p_y}
\end{equation}
is an example of the polyhomeostatic principle. Minimizing
$F[p]$ corresponds to optimizing a given function $f(p)$ of
the neural activity distribution $p(y)$. The resulting 
adaption rates will then influence the timeline $y(t)$ 
of the neural activity. This is an example of guided
self-organization, since the target functional is
expressed in terms of general statistical properties of
the dynamical flow, independently of an eventual semantic
content. The explicit form and derivation of the adaption 
rates will be discussed further below, both for the 
polyhomeostatic functional (\ref{eq_functional_poly}) 
and for the energy functional (\ref{eq_functional_E}).

\section{Equations of Motion}
\label{section_Equations_of_Motion}

There are several venues for deriving equations
of motions from a given target functional. One
uses variational calculus, within classical mechanics,
when deriving the Lagrange equations of motion. In
classical mechanics the target functional, the action,
needs to be stationary with respect to an arbitrary
variation of the trajectory. Here we will consider
instead generic objective functions which are to
be minimized.

Minimizing an objective function is a very generic task
for which a wide range of methods and algorithms
have been developed
\cite{papadimitriou1998combinatorial,goldberg1989genetic,kennedy1995particle}.
Here we are however interested in a different aspect.
Our aim is not to actually find the global minimum of a given 
objective function, or any stationary point, which is
not of interest per se. Objective functions serve
as a guiding principle and equations of motion induced
by minimizing a given objective function will tend to minimize
it. Other driving influence will however in general compete
with this goal and it is this very competition which may result 
in complex and novel dynamical states.

For an objective function which is an explicit function
of the dynamical variable, like the energy functional
(\ref{eq_functional_E}), the equations of motion just
correspond to the downhill flow within the energy 
landscape,
\begin{equation}
\dot x_j \ =\ -\frac{1}{T_e}\,\frac{\partial}{\partial x_j}
E(\{x_k\}) ~, 
\label{eq_EoM_energy}
\end{equation}
where the timescale $T_e$ of the flow in normally set to
unity, $T_e=1$. In our case we obtain
\begin{equation}
\dot x_k \ =\ -\Gamma x_k \,+\,
a_k y_k(1-y_k)\sum_j w_{kj} y_j~,
\label{eq_dot_x_Hopfield}
\end{equation}
where we have used (\ref{eq_sigmoidal}) and
\begin{equation}
y'(x)\ =\ {\partial y\over \partial x}\ =\ a y (1-y)~.
\label{eq_y_prime}
\end{equation}
The dynamical system (\ref{eq_dot_x_Hopfield})
just corresponds to a network of leaky integrators
\cite{hopfield1982neural,hopfield1984neurons},
with the $x_k$ and $y_k$ corresponding to the 
membrane potential and the mean neural firing rate
respectively. The neurons are coupled through
the weight matrix $w_{kj}$, the synaptic weights.
The term $a_k y_k(1-y_k)$ in front of the 
inter-neural coupling is present only when deriving
(\ref{eq_dot_x_Hopfield}) from the energy functional
\cite{linkerhand2012generating}, and not when 
formulating equivalent neural updating rules directly
from neurobiological considerations \cite{olshausen1993neurobiological}.

The polyhomeostatic functional (\ref{eq_functional_poly})
is used to derive adaption rules for the intrinsic parameters
$a_i$ and $b_i$ of the transfer function
(\ref{eq_sigmoidal}). The lack of an explicit 
dependence on either $a_i$ or $b_i$
rules out adaption rules like
$\dot a_i \propto - \partial F[p]/\partial a_i$,
which would be the equivalent to (\ref{eq_EoM_energy}).
It is however possible to derive implicit adaption
rules, for which the minimization of the objective
functions $F[p]$ is performed stochastically in the
sense that the time-averaged firing rate $p(y)$
is sampled along the flow during the time evolution.
For this purpose we change variables and rewrite
the generating functional
\begin{equation}
F[p] \ =\ \int p(x) f\big(p(y)/y')\big) dx,
\qquad\quad
p(y)dy = p(x)dx
\label{eq_functional_poly_x}
\end{equation}
as an expectation value over the distribution $p(x)$
of the membrane potential $x$, the input. The
transfer function (\ref{eq_sigmoidal}) maps the
input of a neuron to its output and adaption rules
for the intrinsic parameters should hence not 
depend explicitly on the actual distribution $p(x)$
of the input, they should be universal in the sense
that the intrinsic adaption rules should abstract
from the actual semantic content of the information
being processed. Noting that $p(x)$ does not depend
explicitly on the gain $a$ and the threshold $b$,
we have
\begin{equation}
\frac{\partial}{\partial\theta} F[p] \ =\ 
\int dx\, p(x) \frac{\partial}{\partial\theta}f\big(p(y)/y')\big),
\qquad\quad \theta=a,b~.
\label{eq_partial_theta_F}
\end{equation}
For the overall global minimum of $F[p]$ the weighting
with respect to the input distribution $p(x)$ would
be needed to be taken into account. As we are however
interested only in adaption rules abstracting from
the actual form of the input distribution, and noting 
that $p(x)\ge 0$ is positive definite, we demand that the
adaption process should lead to a uniform minimization
of the kernel of (\ref{eq_partial_theta_F}),
\begin{equation}
\dot\theta \ =\ -\epsilon_\theta
\frac{\partial}{\partial\theta}f\big(p(y)/y')\big),
\qquad\quad \theta=a,b~,
\label{eq_dot_ab_f}
\end{equation}
where $\epsilon_\theta$ are the respective adaption rates.
The adaption process should generally be slow, as typical
for metalearning, and the adaption rates $\epsilon_a$ and
$\epsilon_b$ small. In this case the updating rule
(\ref{eq_dot_ab_f}) will statistically sample the input 
distribution $p(x)$, as an average over time, and become 
equivalent with (\ref{eq_partial_theta_F}).

The adaption rates (\ref{eq_dot_ab_f}) are generic and
need to be concretized for a specific polyhomeostatic function
$f(p)$. A straightforward target functional for the problem
of allocating time is to consider a target distribution
function $q(y)$ for the neural firing rate. In this case
the functional
\begin{equation}
F[p] \ =\ \int p(y) f(p(y)) dy,
\qquad\quad
f(p) = \ln(p/q)
\label{eq_F_p_KL}
\end{equation}
corresponds to the Kullback-Leibler divergence
\cite{gros2008complex}, which is a positive 
definite measure for the similarity of two distribution 
functions $p$ and $q$. The Kullback-Leibler divergence
is minimal whenever $p(y)$ and $q(y)$ are as similar 
as possible, within the configuration of all dynamically
realizable firing rate distributions $p(y)$.

\begin{figure}[t]
\centerline{
\includegraphics[width=0.25\textwidth,angle=270]{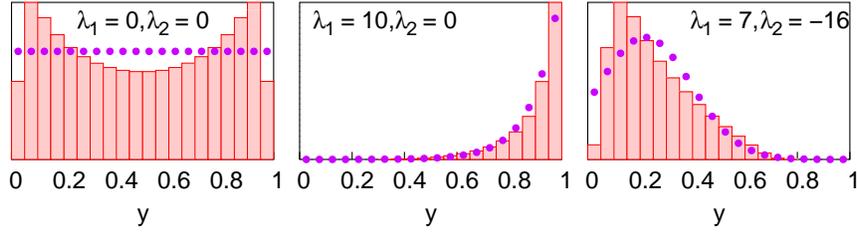}
           }
\caption{The results of the intrinsic adaption rules 
(\ref{eq_Delta_a}) and (\ref{eq_Delta_b}) for the time
averaged firing rate distribution (boxes, see Eq.~(\ref{eq_p_y}))
of a single neuron driven by a white-noise input and for
several information maximizing target distributions
(points, see Eq.~(\ref{eq_entropy_q})).
        }
\label{fig_singleNeuronAdaption}
\end{figure}

The target firing rate distribution $q(y)$ could be any
positive and normalized distribution function.
Here we demand that $q(y)$ should maximize
Shannon's information entropy $-q\ln(q)$, which
can be achieved using variational calculus:
\begin{equation}
0 \ =\ -\delta \int q\left[\ln(q)-\lambda_1 y-\lambda_2y^2\right]dy,
\qquad\quad
q(y) \propto \mathrm{e}^{\lambda_1y+\lambda_2y^2}~.
\label{eq_entropy_q}
\end{equation}
Here $\lambda_1$/$\lambda_2$ are suitable Lagrange
parameters enforcing a given mean/variance. The flat
distribution $\lambda_1=\lambda_2=0$ maximizes information
entropy in the absence of any constraint.
Using (\ref{eq_dot_ab_f}) and $y'=ay(1-y)$, see
Eq.~(\ref{eq_y_prime}), we obtain then the adaption 
rules \cite{triesch2005gradient,triesch2007synergies,markovic2010self,linkerhand2012self}
\begin{eqnarray}
\label{eq_Delta_a}
\dot a &=& \epsilon_a
    \left({1\over a}+(x-b)\,\Delta\tilde\theta\right) \\
\dot b &=& \epsilon_b\, (-a)\,\Delta\tilde\theta,
\qquad\quad
\Delta\tilde\theta \ = \ (1-2y) \,+\, y(1-y)
    \left[\lambda_1+2\lambda_2y\right]~.
\label{eq_Delta_b}
\end{eqnarray}
In Fig.~\ref{fig_singleNeuronAdaption} we present the results
for a single polyhomeostatically adapting neuron, driven
by white noise, for various target distributions $q(y)$.
Note that there are only two intrinsic parameters,
the threshold $b$ and the gain $a$, to be optimized
and that the transfer function (\ref{eq_sigmoidal}) can
hence not change, during the adaption process,
its functional form arbitrarily.  The firing 
rate distribution $p(y)$ approximates, considering 
this limitation, the target distribution $q(y)$ 
remarkably well, an exemplification
of the principle of targeted self-organization.

\begin{figure}[t]
\centerline{
\includegraphics[width=0.75\textwidth,angle=0]{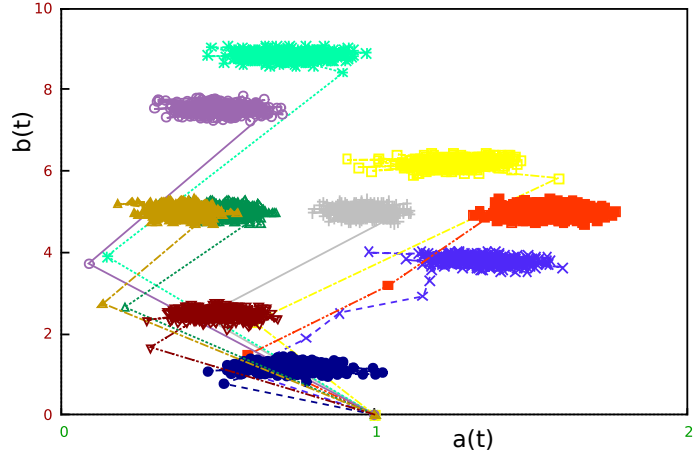}
           }
\caption{Sample trajectories $(a(t),b(t))$ resulting
from the intrinsic adaption rules 
(\ref{eq_Delta_a}) and (\ref{eq_Delta_b}), color coded
for various parameters $\lambda_1$ and $\lambda_2$ 
of the target distributions $q(t)$, compare
Eq.~(\ref{eq_entropy_q}). All trajectories start
at $(a(0),b(0))=(1,0)$ and then settle into distinct
regions of phase space, where they perform a confined
stochastic walk, due to the white-noise input.
        }
\label{fig_singleNeuronAdaptionTime}
\end{figure}

\section{Adaptive Phase Space}
\label{section_Adaptive_Phase_Space}

It is illuminating to investigate somewhat in detail
the behavior of the adaption process in the phase space 
$(a,b)$ of the intrinsic adaption parameters, and
to study individual trajectories $(a(t),b(t))$.
In Fig.~\ref{fig_singleNeuronAdaptionTime} we present
a selection of trajectories for distinct realizations
of the target distribution $q(y)$, as given by
Eq.~(\ref{eq_entropy_q}). The neuron is driven by a 
white noise input, the starting gain and threshold
are $a=1$ and $b=0$, for all trajectories. After a
relatively fast initial transient the intrinsic
parameters settle to distinct respective regions
in the phase space, where they perform a stochastic motion,
reflecting the white-noise character of the driving
input. Three of the resulting firing rate distributions
$p(y)$ are shown in Fig.~\ref{fig_singleNeuronAdaption}.

\begin{figure}[t]
\centerline{
\includegraphics[width=0.75\textwidth,angle=0]{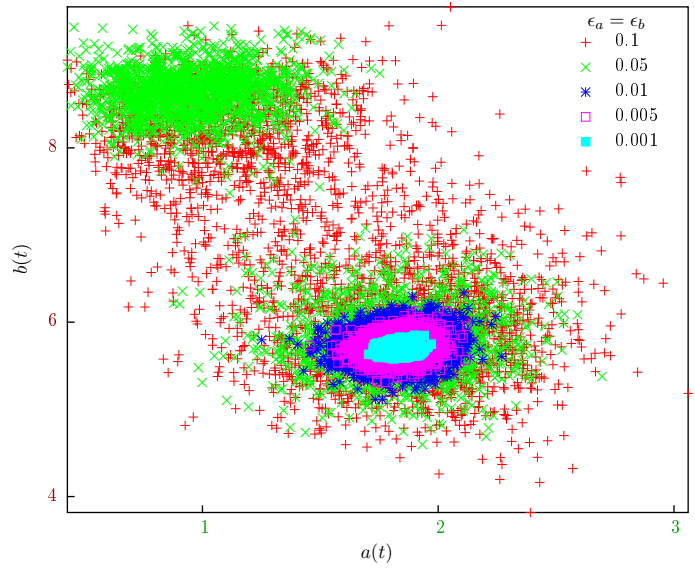}
           }
\caption{Sample trajectories $(a(t),b(t))$ resulting
from the intrinsic adaption rules 
(\ref{eq_Delta_a}) and (\ref{eq_Delta_b}), color coded
for various adaption parameters
$\epsilon_a=\epsilon_b$, as given in the legend.
The single neuron is driven by white noise and
the target distribution, see Eq.~(\ref{eq_entropy_q})
is bimodal, parameterized by $\lambda_1=-20$ and $\lambda_2=18.5$.
For moderate large adaption rates the system is able to
make stochastically driven tipping transitions between
two local minima \cite{linkerhand2012self}.
        }
\label{fig_stochasticTipping}
\end{figure}

The target distribution $q(y)$, see Eq.~(\ref{eq_entropy_q}),
can be selected to be bimodal, which is generally the case
for inverse Gaussians having $\lambda_1<0$ and $\lambda_2>0$.
In Fig.~\ref{fig_stochasticTipping} we present the adaptive
walk through phase space $(a(t),b(t))$ for a bimodal target
distribution having $\lambda_1=-20$ and $\lambda_2=18.5$ and
for various adaption rates $\epsilon_a=\epsilon_b$.
When the adaption process is very slow, viz for small
$\epsilon_a$ and $\epsilon_b$ the system average over
extended periods of the stochastic input and the dynamics
becomes smooth \cite{linkerhand2012self}, fluctuating
with a reduced amplitude around a certain target region 
in phase space, just as illustrated in
Fig.~\ref{fig_singleNeuronAdaptionTime}.

For a bimodal target distribution $q(y)$ there may however
be two local minima in adaptive space, since the transfer
function (\ref{eq_sigmoidal}) is always monotonic. For
any given pair of intrinsic parameters the system can
hence approximate well only one of the two peaks
of a bimodal transfer function. For small adaption 
it remains trapped in one of the local minima, but larger
adaption rates $\epsilon_a$ and $\epsilon_b$ will lead to
an enhanced sensibility with respect to the stochastic
driving, inducing stochastic tipping transitions between
the two local minima. This is a striking realization
of the principle of guided self-organization.

\section{Self-Organized Dynamical States}
\label{section_Self_Organized_Dynamical_States}

As a second example for the functioning of polyhomeostatic
optimization we consider a network of $N$ randomly 
interconnected neurons,
$$
x_k \ =\ \sum_{j\ne k} w_{kj}\, y_j
$$
which corresponds to (\ref{eq_dot_x_Hopfield}) 
in the anti-adiabatic limit $\Gamma\to\infty$
(and without the factor $y(1-y)$). For the
synaptic weights we select
\begin{equation}
w_{ij} \ =\ \left\{
\begin{array}{rcl}
+1/\sqrt{K} &\quad& \mbox{with probability}\quad \rho_{exc} \\
-1/\sqrt{K} &\quad& \mbox{with probability}\quad 1-\rho_{exc} 
\end{array}
\right. ~,
\label{eq_w_ij_random}
\end{equation}
where $K$ is the in-degree. The system is balanced
for $\rho_{exc}=1/2$. As a second control parameter, 
besides the fraction $\rho_{exc}$ of excitatory links, 
we consider the average target activity $\mu$,
\begin{equation}
\mu \ =\ \int y q(y) dy,
\qquad\quad
\int  q(y) dy=1~,
\label{eq_def_mu}
\end{equation}
which is taken to be uniform, viz identical for all
sites.

\begin{figure}[t]
\centerline{
\includegraphics[width=0.85\textwidth,angle=0]{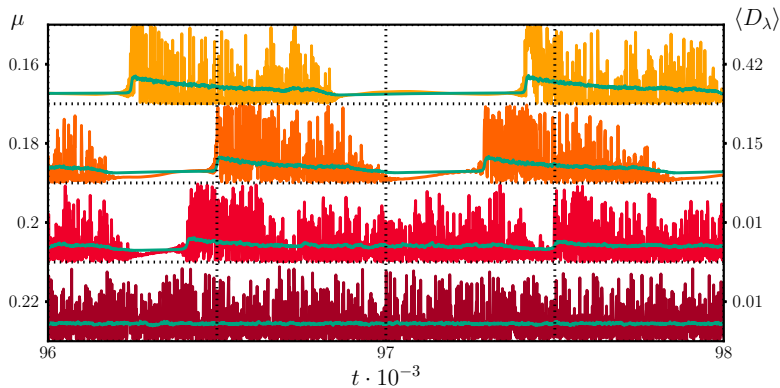}
           }
\caption{For a network of $N=1000$ adapting neurons,
according to Eqs.~(\ref{eq_Delta_a}) and (\ref{eq_Delta_b}), 
the activity of a randomly selected neuron and the average
neural activity (green line). The network is balanced,
with as many excitatory and inhibitory links, randomly selected
according to Eq.~(\ref{eq_w_ij_random}). Shown are results for
various target mean activities $\mu$, see Eq.~(\ref{eq_def_mu}).
The right-hand axis is not a scale, the numbers are the
values of the network-averaged Kullback-Leibler divergence
$\langle D_\lambda\rangle$, as defined by Eq.~(\ref{eq_F_p_KL}).
One observes that the mean target activity $\mu$ entering
the polyhomeostatic generating functional acts as a parameter
controlling the resulting self-organized dynamical state
\cite{markovic2012intrinsic}.
        }
\label{fig_adaptingNetwork_mu}
\end{figure}

In Fig.~\ref{fig_adaptingNetwork_mu} we present
the results for a balanced network with $N=1000$ 
adapting neurons, and an in-degree of $K=100$. 
Shown are both the activity of a single,
randomly selected site and the average activity,
averaged over all sites. We notice that the network
enters distinct dynamical states, as a function 
of the mean target activity $\mu$
\cite{markovic2010self,markovic2012intrinsic}. 
For intermediate target activity levels the dynamics
is chaotic, for smaller mean activities $\mu$ a
regime with intermittent bursts is observed.
One has hence the possibility to tune the
self-organized dynamical state through the
target set by the polyhomeostatic generating 
functional, an example of targeted self-organization.
Interestingly the overall value of the network-averaged
Kullback-Leibler divergence is minimal in the
chaotic state.

\begin{figure}[t]
\centerline{
\includegraphics[width=0.85\textwidth,angle=0]{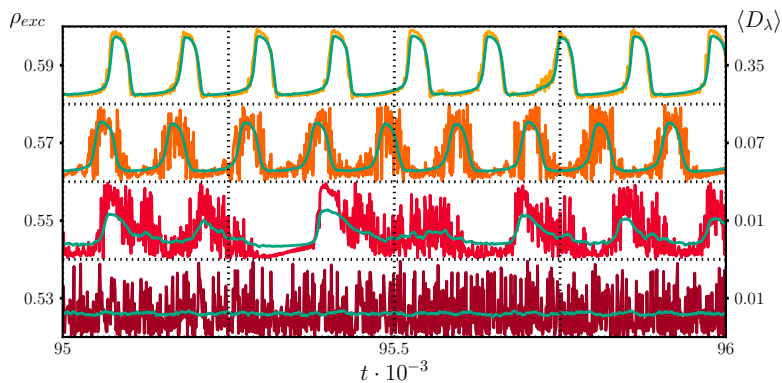}
           }
\caption{For networks containing $N=1000$ adapting 
neurons with an in-degree $K=100$ and a target 
mean activity $\mu=0.3$, see Eq.~(\ref{eq_def_mu}),
the activity of a randomly selected neuron and the average
neural activity (green line). The networks are not balanced,
having a slight excess $\rho_{exc}$ of randomly 
selected excitatory links see Eq.~(\ref{eq_w_ij_random}). 
Also given (on the right) are the respective values
of the network-averaged Kullback-Leibler divergence
$\langle D_\lambda\rangle$, as defined by Eq.~(\ref{eq_F_p_KL}).
The network shows a transition between chaos and synchronization,
as a function of $\rho_{exc}$ \cite{markovic2012intrinsic}.
        }
\label{fig_adaptingNetwork_exc}
\end{figure}

In Fig.~\ref{fig_adaptingNetwork_exc} we present
the results for the same network of $N=1000$ sites as 
in Fig.~\ref{fig_adaptingNetwork_mu}, but this time
the network is not balanced, $\rho_{exc}>1/2$. The
mean target firing-rate activity is kept constant
at $\mu=0.3$. For larger values of $\rho_{exc}$
the network synchronizes, not surprisingly, as
a result of the predominance of positive feedback
loops. For values of $\rho_{exc}$ close to the balanced
state, the system is chaotic, with a large variability
around a partly synchronized state in between.
One can regard $\rho_{exc}$ as a controlling parameter
of the energy functional (\ref{eq_functional_E}),
which hence allows to guide the self-organization of
the resulting dynamical state. The value of the 
Kullback-Leibler divergence is, again, lowest in the
chaotic state, which explores phase space best.

\section{Discussion}
\label{section_Discussion}

A self-organizing process may be guided
by presenting to the system one or more targets.
If these targets are very concrete they may
destroy the self-organizing character of the process,
resulting in a driving force. One possibility
to achieve a gentle way of controlling a self-organizing
process is to formulate the targets in terms of
statistical properties of the desired dynamical state,
with a basic example being the time-average distribution
function of activities. Optimizing the distribution of
activities is an example of a time-allocation problem,
which is intrinsically of polyhomeostatic nature.

A given set of goals may be achieved be a range of different
tools, for example using evolutionary algorithms. In this
treatise we have discussed the perspective, together
with concrete examples, of explicitly deriving equations 
of motions from generating functionals incorporating
polyhomeostatic and other targets. We believe that this
approach offers several advantages. Having explicit
time evolution equation at hand is, in our view, mandatory
for time-efficient simulations and applications.
Generating functionals can furthermore be seen as a route
for solving the control problem, as they offer a substantial
dimensional reduction in the number of free parameters.
This is a particularily attractive feature, in view of the
raising appreciation that the neuromodulator control
system in the brain tunes the relative stability of 
a wide range of possible dynamical operative states 
of the affected downstream circuits.

From an alternative perspective one may view generating
functionals also as a middleway between the study
of simplified model systems and biological realistic
simulations.
\begin{center}
\colorBox{
  \parbox{0.20\textwidth}{
     \centerline{simple}
     \centerline{model systems}
        }} $\longrightarrow$
\colorBox{
  \parbox{0.20\textwidth}{
     \centerline{generating}
     \centerline{functionals}
        }} $\longrightarrow$
\colorBox{
  \parbox{0.20\textwidth}{
     \centerline{detailed\,/\,realistic}
     \centerline{simulations}
        }}
\end{center}
Model systems may constitute important reference models,
for understanding and developing key concepts and
methods. Detailed simulations are, at the other extreme,
often indispensable for obtaining a realistic comparison with
experimental data, having however the drawback that
an in-depth understanding is in general not achievable.
We propose generating functionals as a venue for
building increasingly complex dynamical systems and
cognitive architectures, a venue which allows for the
control of the operating modi of the system by tuning
a limited number of high-level control parameters 
incorporating the targets of the respective 
generating functionals.

\bibliographystyle{apalike}






\backmatter
\printindex


\end{document}